\documentclass[12pt]{article}
\usepackage{glas}
\def\mb#1{\mbox{\small {#1}}}
\def\tb#1{\mbox{\tiny {#1}}}
\def\oos{${\cal O}(\alpha \alpha_s^2)$}
\def\os{${\cal O}(\alpha \alpha_s)$}
\begin{document}
\begin{titlepage}{GLAS-PPE/1999--18}{November 1999}
\title{Hard Photoproduction \\ and the Structure of the Photon}

\author{Laurel Sinclair \\ for the ZEUS and H1 Collaborations}

\begin{abstract}
A pedagogical introduction to the experimental results
on hard photoproduction at HERA is provided.  Then the latest 
results in this field from ZEUS and H1 are reviewed.
\end{abstract}

\vfill
\conference{Invited talk at the Ringberg workshop:``New Trends in HERA Physics''
\\
Tegernsee, Germany, 1999}

\end{titlepage}

\section{Introduction}

It has now been firmly established that the photon can interact strongly, as 
though it were a hadron.  Indeed, two of the first physics results to be 
published  by the HERA experiments H1 and ZEUS were the measurements of 
the total photoproduction cross section~\cite{ZEUS_totgp,H1_totgp}.
These confirmed the cross section of the photon to be of order 100~$\mu$b,
close to the area of a typical hadron.

Of course the total cross section is dominated by
peripheral collisions which cannot be described in perturbative QCD.
In the presence of a hard scale the photon proton cross section factorizes
into terms describing the photon and proton structures, and a hard QCD
subprocess.
Then a perturbative expansion may be applied to determine the subprocess
cross section (for an introduction to the theory of hard photoproduction
see Michael Klasen's contribution to this proceedings).

With subsequent data sets H1 and ZEUS measured the photoproduction
cross section for events in the perturbatively calculable regime, i.e.
with jets of at least
$E_T^{\mb{jet}} > 5$~GeV~\cite{H1_early,ZEUS_early}.
This cross section is naturally much
smaller, of the order of 10~nb, however the hadronic nature of the photon 
is still prevalent.
The HERA experiments clearly established two classes of contribution to the
photoproduction of jets: the direct process
in which the photon itself participates in the hard subprocess
and the resolved process in which the photon fluctuates into a hadronic 
object and one of its partonic constituents participates in the hard
subprocess.
For instance, H1 found an excess of energy in the rear direction, over what
was expected for direct photoproduction processes~\cite{H1_eflow}.  This 
energy could be attributed to the presence of a photon remnant jet in the
resolved photon process.

\begin{figure}[htb]
\begin{center}
\includegraphics[width=1.\textwidth]{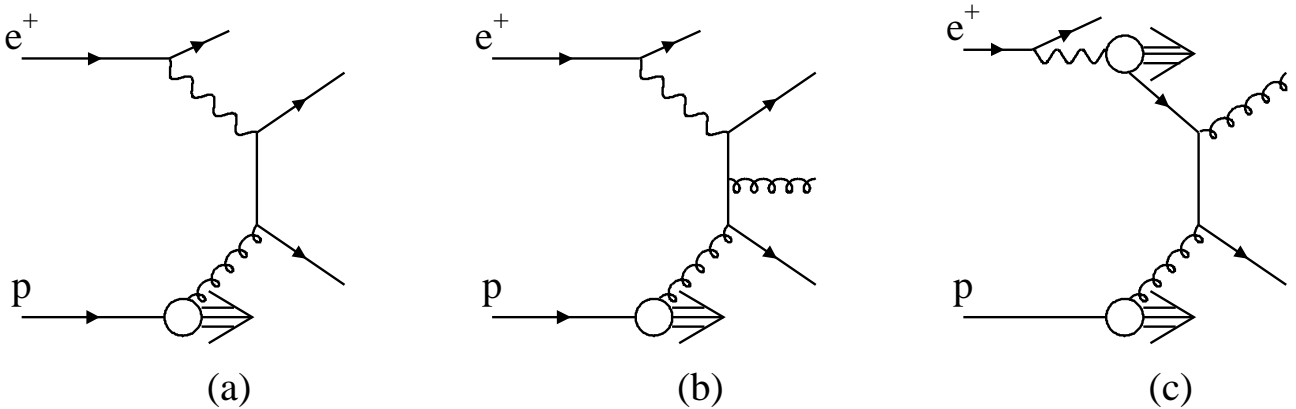}
\end{center}
\caption[]{Illustration of diagrams for (\textbf{a}) LO direct 
(\textbf{b})
NLO direct and 
(\textbf{c}) LO resolved photoproduction processes}
\label{fig:dirres}
\end{figure}

An unambiguous distinction between resolved and direct processes exists
only at leading order (LO).  In Figures~\ref{fig:dirres}(a) and (b) examples of 
direct diagrams in LO and next-to-leading order (NLO), respectively, are
shown.  Figure~\ref{fig:dirres}(c) shows an example of a resolved diagram at 
leading order.
Clearly, if the outgoing quark line in the NLO direct
diagram~\ref{fig:dirres}(b) has small transverse momentum then this process could as well be 
represented by the LO resolved diagram~\ref{fig:dirres}(c).  Therefore some 
prescription must be introduced in order to make a well-defined distinction
between direct and resolved processes.

To this end the observable quantity $x_{\gamma}^{\mb{OBS}}$ has been defined:
\begin{displaymath}
x_{\gamma}^{\mb{OBS}} \equiv \frac
{\sum_{\mb{jets}}E_T^{\mb{jet}}e^{-\eta^{\tb{jet}}}} {2yE_e},
\end{displaymath}
where $E_T^{\mb{jet}}$ and $\eta^{\mb{jet}}$ are the jet transverse energy
and pseudorapidity respectively, $y$ is the fraction of the electron's energy
carried by the incoming photon, $E_e$ is the incoming electron energy, and
the sum runs over the two highest  $E_T^{\mb{jet}}$ jets within the
$\eta^{\mb{jet}}$ acceptance.
At leading order $x_{\gamma}^{\mb{OBS}}=1$ for direct processes and
$x_{\gamma}^{\mb{OBS}}<1$ for resolved processes.
However, $x_{\gamma}^{\mb{OBS}}$ is well defined theoretically to any order of 
perturbation theory (provided, of course, that the jet-finding algorithm
is well-behaved).  Therefore, a quantitative
confrontation of a measurement of a ``direct'' or 
``resolved'' photoproduction
cross section with a pQCD calculation can be made provided the separation 
between the direct and resolved regions is made in terms of 
$x_{\gamma}^{\mb{OBS}}$.

\begin{figure}[htb]
\begin{center}
\includegraphics[width=.7\textwidth]{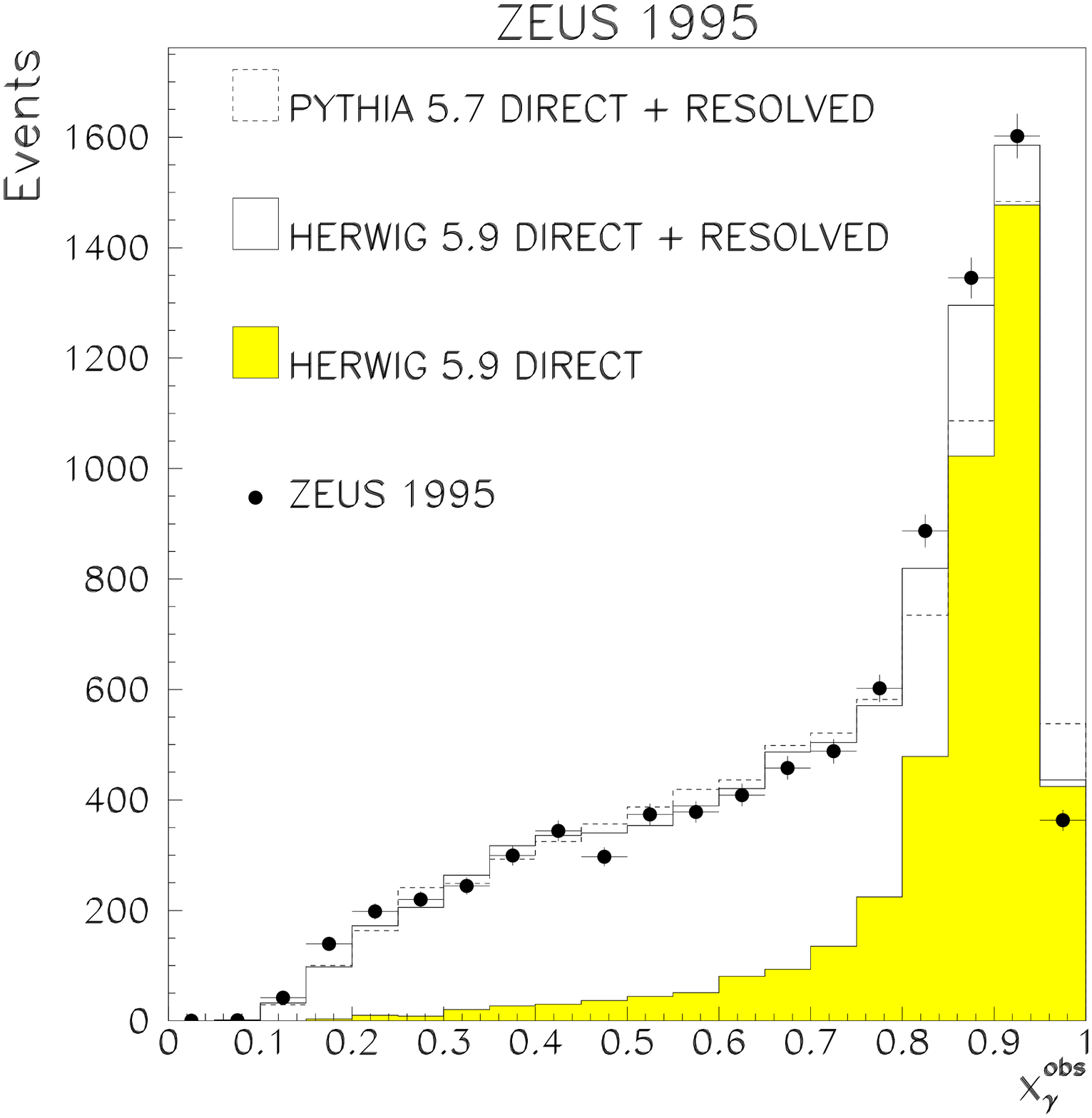}
\end{center}
\caption[]{The $x_\gamma^{\tb{OBS}}$ spectrum for events with 
$E_T^{\tb{jet1, jet2}} > 11, 14$~GeV,
compared to the HERWIG 5.9 and the PYTHIA 5.7 Monte Carlo predictions.
The direct component from the HERWIG Monte Carlo is shown separately as the 
shaded histogram. Only statistical uncertainties are plotted
\label{fig:ZEUS_xgam}}
\end{figure}

In Figure~\ref{fig:ZEUS_xgam} a measured $x_{\gamma}^{\mb{OBS}}$ distribution
is shown for photoproduction events containing two jets of
$E_T^{\mb{jet1, jet2}} > 11, 14$~GeV within
$-1 < \eta^{\mb{jet}} < 2$ in the HERA lab frame~\cite{ZEUS_dijet}.
The data show a peak at high $x_{\gamma}^{\mb{OBS}}$ values with a broad
tail extending to $x_{\gamma}^{\mb{OBS}} \sim 0$.  The data are compared
with the predictions of two parton shower Monte Carlo programs,
HERWIG~5.9~\cite{herwig1,herwig2} and PYTHIA~5.7~\cite{pythia1,pythia2}.
The Monte Carlo calculations
implement the QCD matrix elements at leading order only.  The effect of
higher order processes is approximated through initial and final-state
parton showers.  The programs differ in the choice of evolution variable for
the parton shower calculation and also in the technique chosen to convert
the final partonic configuration into colourless hadrons.  Both are able
to provide a good description of the $x_{\gamma}^{\mb{OBS}}$ distribution.
The distribution of the HERWIG events with LO direct
photoproduction subprocesses is shown separately as the shaded histogram.  The
parton showering and hadronizaton phases smear the $x_{\gamma}^{\mb{OBS}}$
values such that this distribution is peaked just below one and can extend to
the lowest available $x_{\gamma}^{\mb{OBS}}$ values.  Nevertheless it is
plain to see that a sample of events with $x_{\gamma}^{\mb{OBS}} > 0.75$ is
essentially of the direct photoproduction type.
ZEUS and H1 have published
several papers in which direct and resolved photoproduction regions are 
defined based on
the $x_{\gamma}^{\mb{OBS}}$ 
observable~\cite{ZEUS_early,ZEUS_dijet} and~\cite{ZEUS_dij95}
to~\cite{ZEUS_dstar} 

\begin{figure}[htb]
\begin{center}
\includegraphics[width=1.\textwidth]{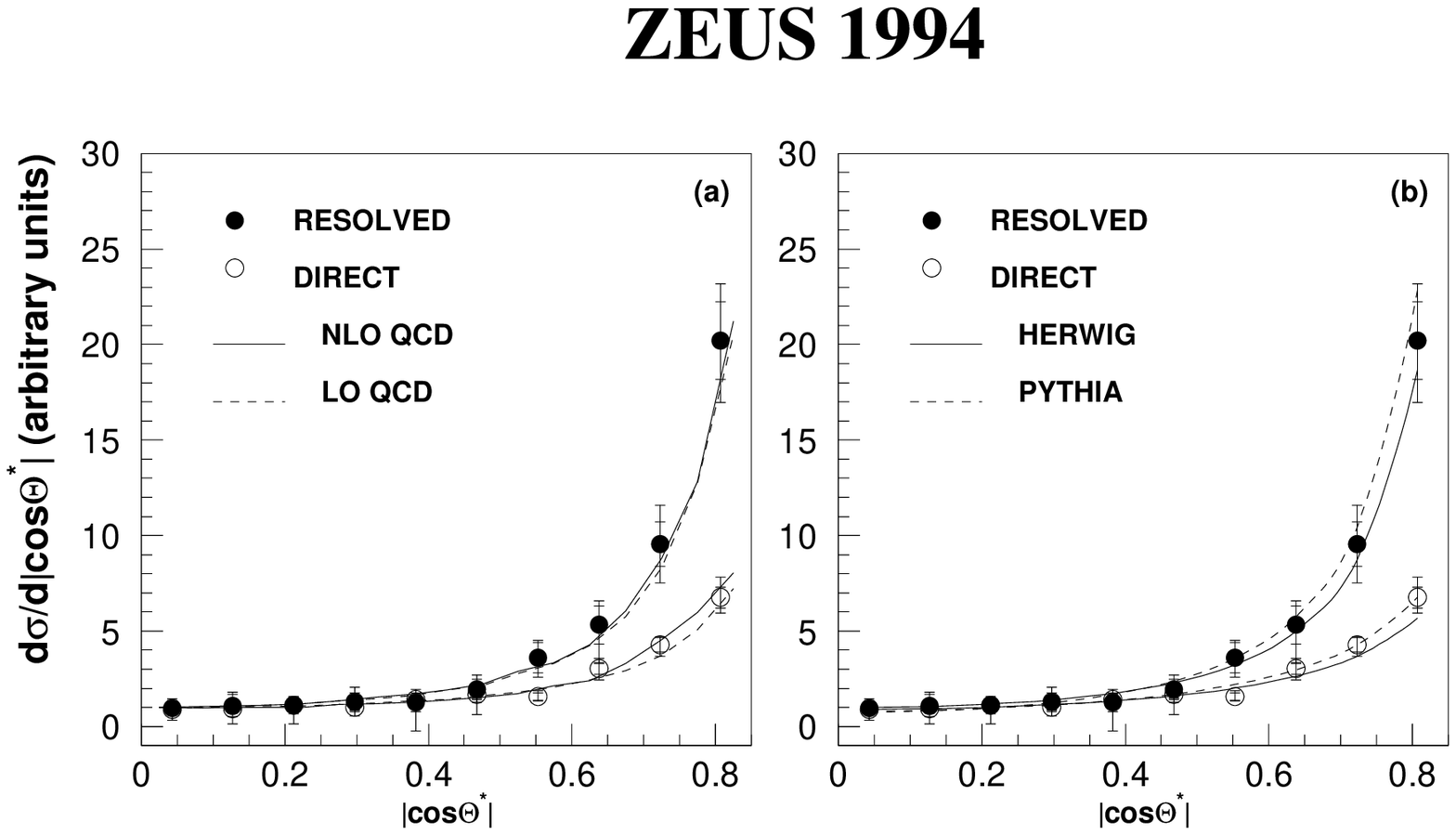}
\end{center}
\caption[]{
$d\sigma/d|\cos\vartheta^*|$ normalized to one at $\cos\vartheta^* = 0$
for $x_{\gamma}^{\tb{OBS}}< 0.75$ ({\it black dots}) and 
$x_{\gamma}^{\tb{OBS}}\ge 0.75$ ({\it open circles}) photoproduction.
In (\textbf{a}), the ZEUS data are compared to NLO predictions
({\it solid line}) and LO predictions ({\it broken line}). 
In (\textbf{b}), the broken line is the PYTHIA distribution and
solid line is the HERWIG distribution. The inner error bars are the 
statistical errors,
the outer error bars are the sum in quadrature of the statistical and 
systematic uncertainties
}
\label{fig:ZEUS_angle}
\end{figure}

An important confirmation of the presence of the direct and resolved
contributions to photoproduction was provided by the ZEUS measurement of
dijet scattering angles~\cite{ZEUS_angles}.
A two jet final state can be completely specified in its centre-of-mass
frame (up to an arbitrary azimuthal rotation) by the dijet invariant mass,
$M_{\mb{2J}}$, and the dijet scattering angle, $\vartheta^*$.
Direct photoproduction processes should proceed
predominantly through quark exchange (the diagram shown at leading order
in Figure~\ref{fig:dirres}(a)).  As the exchanged parton is of spin $1/2$, at leading order
the dijet scattering angle is distributed according to
$1/(1-|\cos\vartheta^*|)$.  In contrast, resolved processes 
are dominated by the exchange of the integer spin gluon
which gives rise to the steeper angular dependence,
$1/(1-|\cos\vartheta^*|)^2$.  ZEUS observed
a steeper angular dependence for the events with $x_{\gamma}^{\mb{OBS}}<0.75$
than for those with $x_{\gamma}^{\mb{OBS}}\ge 0.75$, as shown in 
Figure~\ref{fig:ZEUS_angle}.

This is a compelling observation.  An invariant mass cut of
$M_{\mb{2J}}>23$~GeV has been applied to ensure that the
$E_T^{\mb{jet}} > 6$~GeV requirement does not bias the angular distribution.
Moreover, the dijet scattering angle is defined in the centre-of-mass frame of
the two jets so the
fact that the lower $x_{\gamma}^{\mb{OBS}}$ events are boosted more in
the proton direction 
is not responsible for the differences
observed in the $\cos\vartheta^*$ distributions.
The different $\cos \vartheta^*$ distributions for the high
and low $x_{\gamma}^{\mb{OBS}}$ samples are therefore 
an unambiguous demonstration of the differing
underlying QCD subprocess dynamics.  There is a greater
contribution from gluon exchange processes contributing to the
$x_{\gamma}^{\mb{OBS}}< 0.75$ sample, than there is contributing to the
$x_{\gamma}^{\mb{OBS}}\ge 0.75$ sample.  This is consistent with the
expectation that more resolved photon processes contribute to the
$x_{\gamma}^{\mb{OBS}}< 0.75$ sample.

Thus it has been established that both direct and resolved processes
contribute to photoproduction at HERA.  The study of these processes is now
providing a fruitful forum for the investigation of strong interactions.
Photoproduction processes access the physics involved in the structures
of the photon and proton, in the dynamics of hard subprocesses, and in
the fragmentation and hadronization of the final state partons.

The structure of the photon is probed in deep inelastic $e \gamma$
experiments at $e^+ e^-$ colliders in just the same way as the structure of 
the proton is probed in $e p$ interactions at HERA.  These
measurements of the photon's structure function, $F_2^{\gamma}$, constrain 
the quark densities at intermediate probing energy scales.
However, the gluon density is not directly constrained in these experiments,
nor do the quark density constraints extend to the high energy scale
region accessible at HERA.  It is in these two areas that the HERA
experiments have concentrated their efforts.

Phenomenological ans\"{a}tze for the parton densities of the photon exist.
These generally involve the assumption that at low probing virtualities,
the photon undergoes a quantum fluctuation into a vector meson
or an unbound $q \bar{q}$ state and this
provides the photon's structure.
A DGLAP evolution is then invoked in order to evolve the parton densities
to arbitrary scale, using the $e \gamma$ $F_2^\gamma$ data as a constraint.  
Thus, 
studies of photoproduction sensitive to the photon's structure test
fundamental physical assumptions, in addition to providing a means to
investigate the universality of photon structure data obtained through
different processes.  In sections~\ref{sec:real} and \ref{sec:virtual} the 
latest results from H1 and ZEUS pertaining to the structure of the photon 
are presented.

Perturbative QCD governs the behaviour of the partons emerging from the hard
subprocess.  As the distribution of the jets of hadrons in the final
state bears a close correspondence to the distribution of the underlying
partons, measurements may be designed which are sensitive to the subprocess
dynamics.
The differing dijet angular distributions for direct and resolved
photoproduction have already been discussed.  As the available luminosity
delivered by HERA has increased, it has become possible to look for unusual
dynamical signatures in high $E_T^{\mb{jet}}$ dijet processes, and also to
investigate the underlying mechanism of three jet production.  Studies of
the matrix element dynamics are presented in section~\ref{sec:matrix}.

Our knowledge is limited about the physics of hadronization, whereby the partons
resulting from a collision are converted to colourless hadrons by
the inexorable confinement force of QCD.
The hadronization occurs at low momentum transfers where the QCD coupling
is much too strong for a perturbative expansion to be relevant.
However, there are experimental results which have provided some information
about this interesting area of physics.  For instance, a universality of the
hadronization of quarks is supported by the measurements of jet shapes in
$e p$ and $e^+ e^-$ collisions~\cite{ZEUS_shapes}.
Recently, a procedure for measuring jet structure in hadronic
collisions has been proposed which is valid for an all-orders calculation
in perturbative QCD~\cite{Mike_subjets}.
In this way, a well-defined comparison of theory and data can be undertaken
which begins well within the regime where a perturbative approach should be
valid and then approaches the mysterious realm of the very strong hadron
producing force.
A measurement of jet substructure using this algorithm is presented in
Section~\ref{sec:structure}.

\section{Real Photon Structure}
\label{sec:real}

\begin{figure}[htb]
\begin{center}
\includegraphics[width=.7\textwidth]{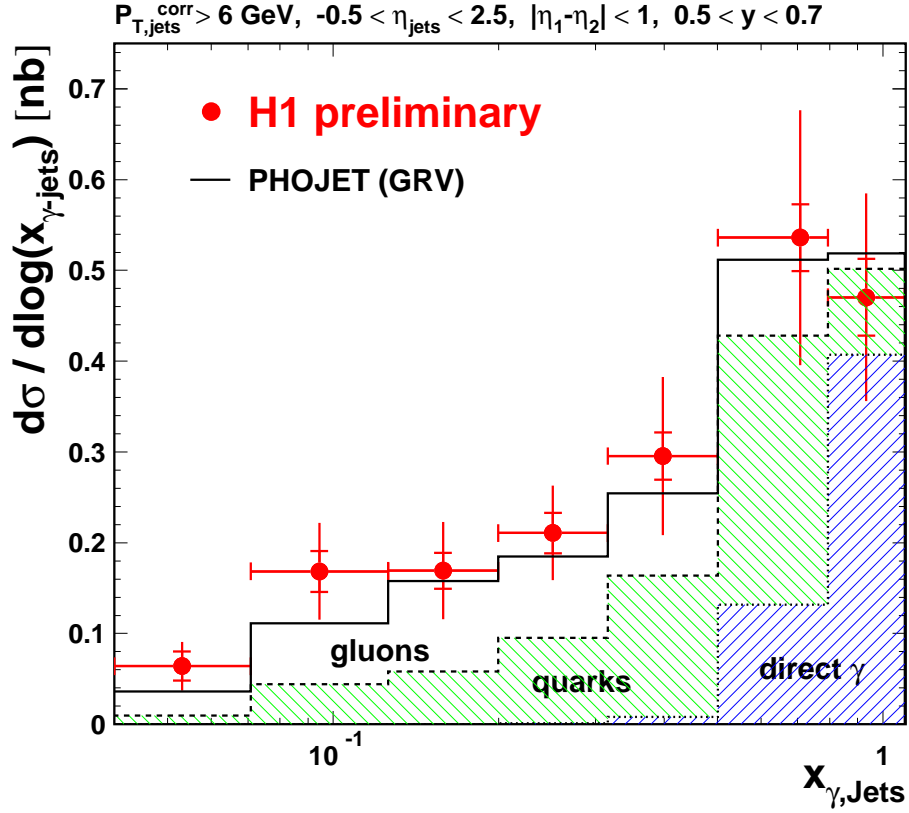}
\end{center}
\caption[]{Dijet cross section in photproduction as a function of 
$x_\gamma^{\tb{jets}}$.  The data are compared to the prediction of
PHOJET.  The direct photon prediction, and the resolved photon predictions
for either quarks are gluons, are shown separately}
\label{fig:H1_xgam}
\end{figure}

A measurement which is sensitive to the gluon density of the photon has been
made by the H1 collaboration~\cite{H1_glue}.  The gluonic component of
the photon dominates at low $x_\gamma$.
Therefore the events must be selected keeping
$E_T^{\mb{jet}}$ as low as possible and allowing $\eta^{\mb{jet}}$ to extend
as far forward into the incoming proton direction as possible.  This is a
difficult kinematic region experimentally, as the energy-scale
uncertainty and angular resolution of the calorimeter are worst here.
There are also theoretical limitations as the contribution from events in
which there is a secondary scatter, and the smearing between a partonic and
hadronic distribution, increase as $E_T^{\mb{jet}}$ lowers.  Nevertheless
it has been possible to make the measurement with sufficient precision to
illustrate the sensitivity of the data to the distribution of gluons in the
photon.
Figure~\ref{fig:H1_xgam} shows
$d\sigma/dx_\gamma^{\mb{jets}}$~\footnote{
$x_\gamma^{\tb{jets}}$ is essentially the same quantity as
has been called $x_\gamma^{\tb{OBS}}$.}
for events
containing two jets of $E_T^{\mb{jet}} > 6$~GeV within
$-0.5 < \eta^{\mb{jet}} < 2.5$.  The data are compared with the predictions
of the leading order plus parton shower Monte Carlo program
PHOJET~\cite{phojet1,phojet2} where the predictions including and 
neglecting the gluons
in the photon are shown separately.  Although there is a large systematic
uncertainty affecting the measurement, within the PHOJET model
a significant gluonic contribution at low $x_\gamma^{\mb{jets}}$
is required by the data
assuming the GRV-LO~\cite{GRVLO1,GRVLO2} parton densities for the photon and proton.

H1 have gone on to subtract the influence of  secondary parton scattering and
unfold the data to a leading order effective parton density, using the
Monte Carlo model.  From this the  modelled quark density has been subtracted 
to yield
the gluon density as a function of $x_{\gamma}$ where now
$x_{\gamma}$ refers to the fraction of the momentum of the photon 
which enters the leading order hard
subprocess.  This is necessarily a model dependent
result; however it can serve to provide insight into
photon structure.  H1 finds that the gluon density rises as
$x_{\gamma}$ decreases.

The ZEUS analysis of photon structure has concentrated on limiting the
data to that kinematic regime in which perturbative QCD should be
applicable, without the need for additional model parameters.  Dijet events
have been selected with
$E_T^{\mb{jet1, jet2}} > 11, 14$~GeV within
$-1 < \eta^{\mb{jet}} < 2$~\cite{ZEUS_dijet}.
The $x_{\gamma}^{\mb{OBS}}$ distribution for this selection is shown 
in Figure~\ref{fig:ZEUS_xgam}.  The PYTHIA and HERWIG predictions which are
compared with the data in this figure contain no simulation of soft underlying
events or of secondary parton scatterings and provide a good description of 
the data.  Also, studies have indicated that hadronization effects in this
kinematic regime should be small, at most around 10\%.
Therefore a strong
interpretation of this data within perturbative QCD can be made.

\begin{figure}[h!]
\begin{center}
\includegraphics[width=.9\textwidth]{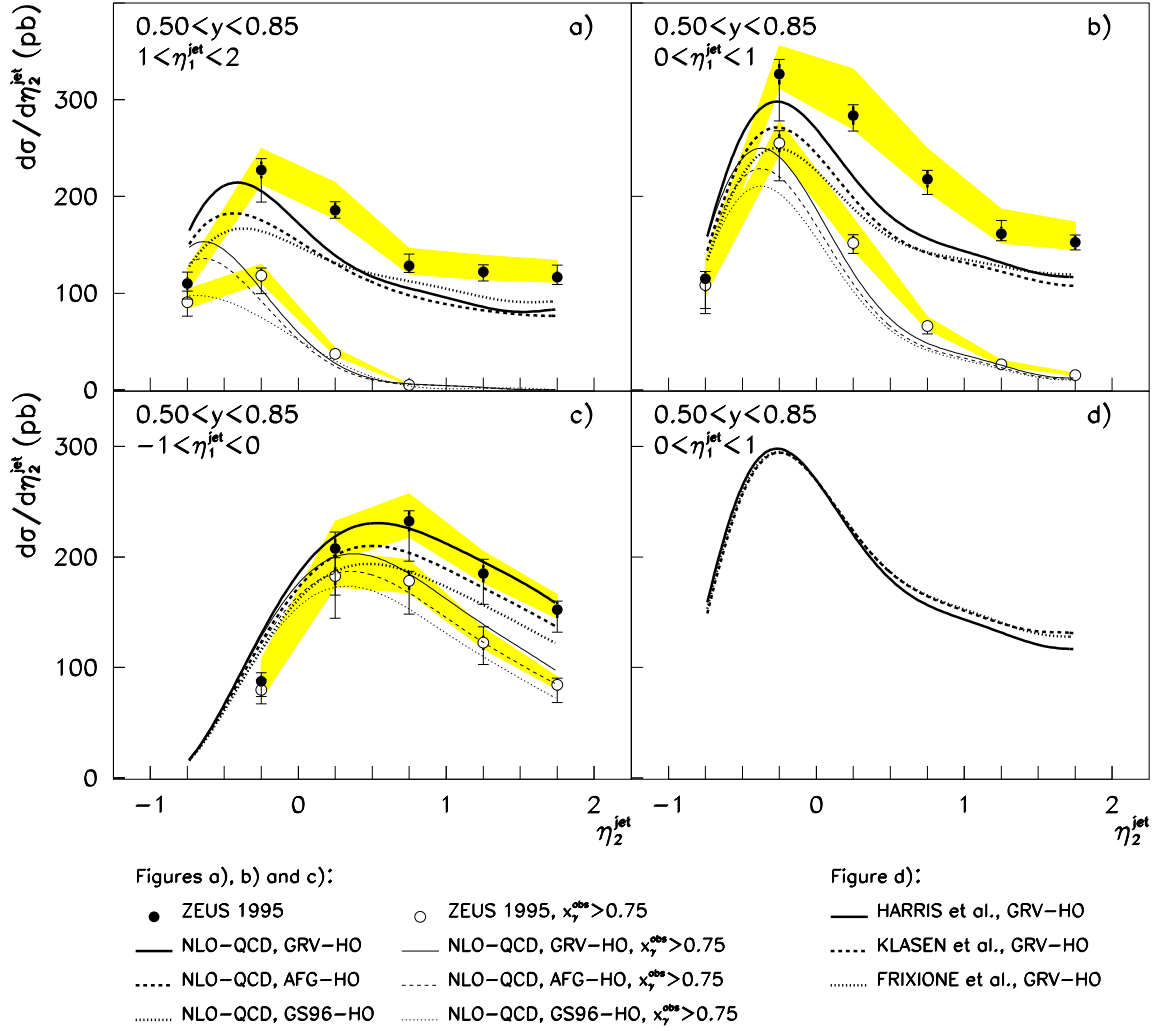}
\end{center}
\caption[]{Figures (\textbf{a}), (\textbf{b}) and 
(\textbf{c}) show the dijet cross section  
as a function of $\eta_2^{\tb{jet}}$ in bins of $\eta_1^{jet}$.
 The filled circles correspond to the entire 
 $x_\gamma^{\tb{OBS}}$ range while the 
open circles correspond to events with $x_\gamma^{\tb{OBS}}> 0.75$. The shaded 
band indicates the uncertainty related to the energy scale. The thick 
error bar indicates the statistical uncertainty and the thin error bar 
indicates the 
systematic and statistical uncertainties added in quadrature. The full, 
dotted and dashed curves correspond to NLO-QCD calculations, using the 
GRV-HO, GS96-HO and the AFG-HO parameterizations for the photon structure, 
respectively. In (\textbf{d}) the NLO-QCD results for the cross section when 
$0<\eta_1^{\tb{jet}}<1$ and for a particular parameterization of the 
photon structure are compared}
\label{fig:ZEUS_highy}
\end{figure}

In Figure~\ref{fig:ZEUS_highy} the cross section for the
$E_T^{\mb{jet1, jet2}} > 11, 14$~GeV selection,
$d\sigma/d\eta_1^{\mb{jet}}$, is shown in bins of $\eta_2^{\mb{jet}}$.  The
kinematic region is restricted to a narrow $y$ range, in order to improve
the sensitivity to the photon's structure.
The cross section is shown separately for $x_{\gamma}^{\mb{OBS}} \ge 0.75$,
indicating that the direct process dominates when the jets tend toward the
incoming photon direction.
NLO perturbative QCD predictions with the three available
photon parton parametrizations~\cite{GRVLO1,grvho1,gs96,afg}
are compared with the data in this figure.
(Note that the calculations have been checked by several
different groups of theorists
as reported in~\cite{Klasen}.)
The predictions underestimate the data in the central rapidity region where
experimental and theoretical uncertainties are expected to be particularly
small.
As previously mentioned, the parton density of the photon is not well
constrained by $e \gamma$ data at these high energy scales.
Therefore it is expected that the parton density of the photon may be
underestimated in this kinematic regime in the currently available parton
density functions of the photon.

In another interesting analysis by H1 the photon remnant jet has been tagged
in low $x_{\gamma}^{\mb{jet}}$ events by running a clustering algorithm,
requiring exactly four jets in the event and defining the photon remnant jet as
that jet closest to the incoming photon direction~\cite{H1_remnant}.  The
$E_T^{\mb{jet}}$ of the remnant jet was found to be correlated with the
$E_T^{\mb{jet}}$ of the highest transverse energy jets.  This points to the
presence of the anomalous component of the photon's structure whereby the
struck parton arises from a $q \bar{q}$ splitting of the photon rather than 
as a constituent of a fluctuation into a vector meson.  The 
``remnant jet'' can then, as
in Figure~\ref{fig:dirres}(b), be viewed instead as one of the outgoing partons
from a next-to-leading order hard subprocess.

Another promising process for constraining the parton density of the
photon is prompt photon production, in which an outgoing quark from the hard
subprocess is balanced not by a gluon, but by a photon (see~\cite{Klasen}
for a complete discussion of the contributing diagrams).
ZEUS has published a measurement of the cross section for prompt photon
production in association with a jet~\cite{ZEUS_promptp}.  A more inclusive
measurement, whereby only the prompt photon is tagged without the jet
requirement, is free of complications due to the matching of the jet
definition in theory and experiment and relatively free of hadronization
corrections.  This then, like high $E_T^{\mb{jet}}$ production, is an area
in which a strong interpretation can be made from the comparision of the
data with the predictions of perturbative QCD.  Of course the cross 
section for prompt photon production is suppressed with respect to jet
production due to the smallness of the electromagnetic coupling constant and
currently the statistics are limited.  Nevertheless the technique of prompt
photon identification has been refined and a first comparison with the
theory indicates a rough agreement~\cite{Klasen}.

\section{Virtual Photon Structure}
\label{sec:virtual}

\begin{figure}[h!]
\begin{center}
\includegraphics[width=.8\textwidth]{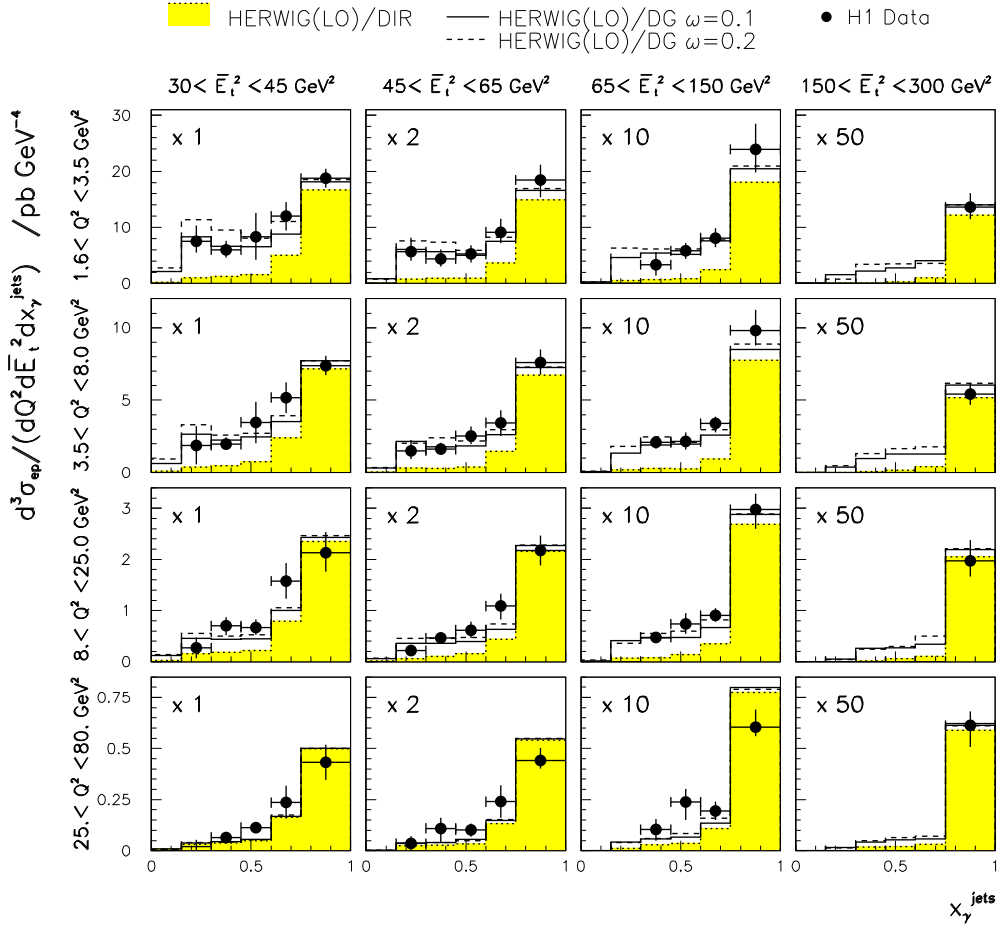}
\end{center}
\caption[]{The differential dijet cross section
$d\sigma_{e p} / (dQ^2 d\bar{E_t}^2dx_{\gamma}^{\tb{jets}})$ shown as a
function of $x_{\gamma}^{\tb{jets}}$ for different regions of $\bar{E_t}$
and $Q^2$.  The scale factors applied to the cross sections are indicated.
The error bar shows the quadratic sum of systematic and statistical errors.
Also shown is the HERWIG model where the direct component is shown as the
shaded histogram}
\label{fig:H1_virt}
\end{figure}

It is expected that as the virtuality of the incoming photon increases, less
time will be available for it to develop a complex hadronic structure.
To test this assumption, H1 has measured the dijet triple differential cross
section,
$d\sigma_{e p} / (dQ^2 d\bar{E_t}^2dx_{\gamma}^{\mb{jets}})$
where $\bar{E_t}$ is the average jet transverse energy and $Q^2$ is the 
negative of the square of the momentum transfer at the scattered lepton 
vertex~\cite{H1_virt}.
The cross section is presented in Figure~\ref{fig:H1_virt} as a function
of $x_{\gamma}^{\mb{jets}}$ in bins of $\bar{E_t}^2$ in the range
$30 < \bar{E_t}^2 < 300$~GeV$^2$ and in bins of $Q^2$ in the range
$1.6 < Q^2 < 25$~GeV$^2$.
The data are concentrated near  $x_{\gamma}^{\mb{jets}} = 1$ with a small
tail to lower values.
Compared with the data are predictions from the HERWIG model, where the
events with a leading order direct subprocess are shown separately by the
shaded histogram.
Looking at fixed $\bar{E_t}^2$, there is clear evidence for the expected
suppression of the resolved processes as $ Q^2$ increases.
However, wherever  $\bar{E_t}^2 \gg Q^2$, the direct processes alone are
insufficient to account for the low $x_{\gamma}^{\mb{jets}}$ events.
Thus there is evidence for resolved photon processes, even well into the deep
inelastic scattering regime, $Q^2>8$~GeV$^2$.

H1 have extended the analysis in their publication~\cite{H1_virt} 
and unfolded the data to
a leading order effective parton density,
$f_{\mb{eff}} \equiv \sum_i^{N_f}(q_i + \bar{q}_i) + \frac{9}{4}g$,
relying on the Monte Carlo simulation to correct the data for hadronization
and higher order effects.
The behaviour of this parton density is consistent with a logarithmic rise
with the probing resolution, $P_t^2$, where $P_t$ is the transverse momentum
of the two outgoing partons at leading order.
This is in contrast with the approximate scaling behaviour which has been
observed for hadrons and reflects the presence of the anomalous, or
$q \bar{q}$ splitting term, unique to the structure of the photon.

The effective parton density is also found to exhibit a dependence on the
photon's virtuality $Q^2$ which is consistent with the expected logarithmic
suppression implemented in the existing virtual photon parton density
functions.
Incorporating an earlier measurement of  $f_{\mb{eff}}$ at $Q^2 = 0$ with 
these data, a drop in  $f_{\mb{eff}}$ with $Q^2$ is indicated.
(To compare the earlier measurement with this one the 
data have been evolved using the GRV~LO
parton densities to the same $P_t^2$ and $x_\gamma$ range.)

The ZEUS collaboration has measured the dijet cross section
$d\sigma/dx_{\gamma}^{\mb{OBS}}$
for jets of
$E_T^{\mb{jet}}> 5.5$~GeV in the three different photon virtuality ranges
$Q^2 \sim 0$ ($Q^2 < 1$~GeV$^2$), $0.1 < Q^2 < 0.55$~GeV$^2$ and
$1.5 < Q^2 < 4.5$~GeV$^2$~\cite{ZEUS_virt}.  This measurement is 
complementary to the H1
analysis in that the $Q^2 \sim 0$ data are analyzed together with the deep
inelastic data at $1.5 < Q^2 < 4.5$~GeV$^2$ in a consistent way.  Also,
making use of an auxiliary small angle electron tagger, the
ZEUS measurement includes data in the important transition region
between photoproduction and deep inelastic scattering,
$0.1 < Q^2 < 0.55$~GeV$^2$.
From the $d\sigma/dx_{\gamma}^{\mb{OBS}}$ measurements, similar conclusions
can be drawn as in the H1 analysis: the population of the
$x_{\gamma}^{\mb{OBS}}$ distribution is suppressed at low values as
$Q^2$ increases yet there is evidence for a resolved component of the photon
even in the deep inelastic scattering regime, $Q^2 > 1.5$~GeV$^2$.

\begin{figure}[htb]
\begin{center}
\includegraphics[width=.7\textwidth]{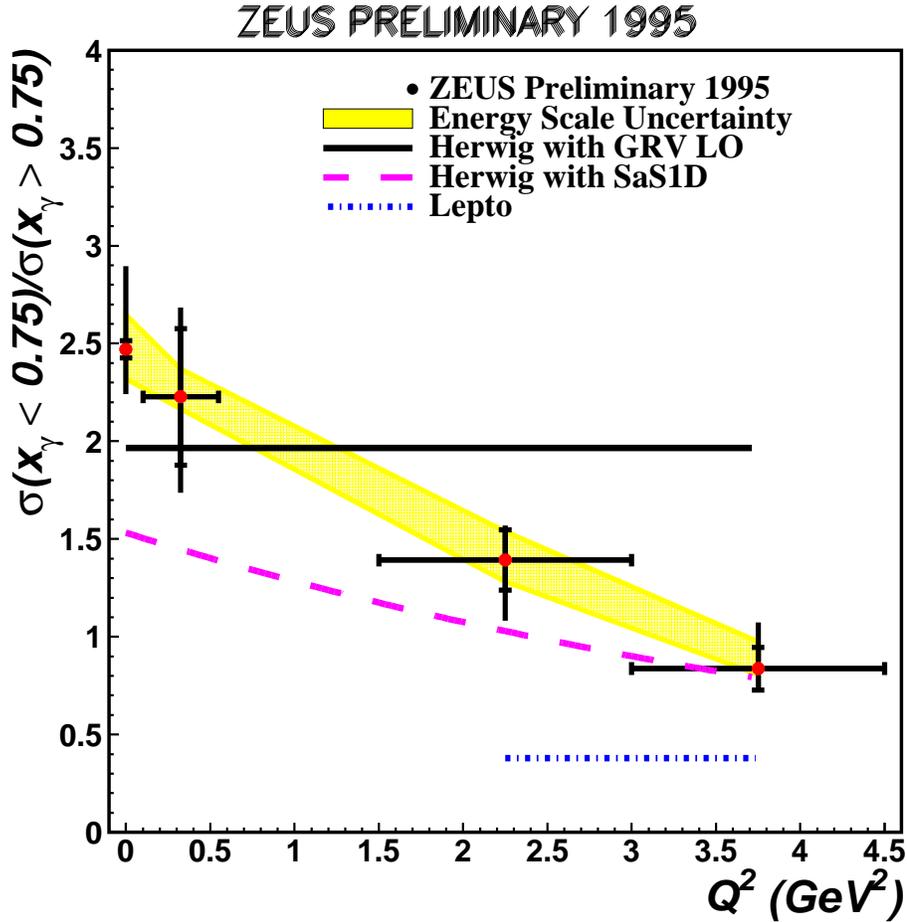}
\end{center}
\caption[]{The ratio of dijet cross sections,
$\sigma(x_{\gamma}^{\tb{OBS}}<0.75)/\sigma(x_{\gamma}^{\tb{OBS}})\ge0.75)$,
as a function of photon virtuality, $Q^2$.  The inner error bar represents
the statistical error and the outer the statistical and systematic
uncertainties added in quadrature.  The band represents the systematic
uncertainty due to the jet energy scale.  Also shown are the predictions
of HERWIG for two different choices for photon parton densities:  GRV
for real photons ({\it full line}) and 
SaS~1D ({\it dashed line}) which includes a
suppression of the photon parton density with increasing photon
virtuality.  The LEPTO predictions are shown for $Q^2>1.5$~GeV$^2$ 
({\it dot-dashed line})}
\label{fig:ZEUS_virt}
\end{figure}

In order to make a precise statement concerning the evolution of the virtual
photon parton densities with $Q^2$, ZEUS has measured the ratio of the
dijet cross section for $x_{\gamma}^{\mb{OBS}} < 0.75$ to that for
$x_{\gamma}^{\mb{OBS}} \ge 0.75$ in bins of $Q^2$~\cite{ZEUS_virt}.
The result is shown in Figure~\ref{fig:ZEUS_virt}.
The ratio of resolved to direct cross sections is found to fall with
$Q^2$.
In comparison, this ratio within the HERWIG model is flat for a photon
parton density which does not fall with $Q^2$ (GRV LO) and 
falling for a  photon parton density which does (SaS 1D~\cite{SaS}).
Therefore the data indicate that the photon parton density is suppressed 
with $Q^2$ in this LO description.
Furthermore, as $Q^2$ increases the data tend toward a leading order 
prediction which does not include any resolved photon contribution
(Lepto~6.5.1~\cite{Lepto}).

\section{QCD Matrix Elements}
\label{sec:matrix}

The influence of the dominant QCD subprocesses on dijet angular distributions
for events with $M_{\mb{2J}} > 23$~GeV has been observed, as already
discussed.
As the integrated luminosity delivered by HERA continues to increase, it
is important to re-measure these distributions in the newly accessible
kinematic regimes, in order to check for new contributing processes to
dijet production.
Using the 1995 to 1997 integrated data set ZEUS has measured the 
$M_{\mb{2J}}$ and $\cos \vartheta^*$ distributions in the kinematic regime 
$M_{\mb{2J}} > 47$~GeV and $|\cos \vartheta^*| < 0.8$~\cite{ZEUS_highmass}.
The measurement extends to a  dijet invariant mass of
$M_{\mb{2J}} \sim 140$~GeV,
and the mass and angular distributions are well described by 
the predictions of perturbative QCD.

For three massless jets, five quantities are necessary to
define the system.  These are defined
in terms of the energies, $E_i$, and momentum three-vectors, 
$\vec{p}_i$, of the jets in the three-jet centre-of-mass frame and 
$\vec{p}_{B}$, the 
beam direction. 
They are the three-jet invariant mass,
 $M_{\mbox{\scriptsize 3J}}$;
the energy-sharing quantities $X_3$ and $X_4$,
$
X_i \equiv {2E_i} / {M_{\mbox{\scriptsize 3J}}};
$
the cosine of the scattering angle of the highest energy jet with
respect to the beam,
$
\cos \vartheta_3 \equiv {\vec{p}_{B} \cdot \vec{p}_3} / 
                        {(|\vec{p}_{B}| |\vec{p}_3|)};
$
and $\psi_3$, the angle between
the plane containing the highest energy jet
and the beam and
the plane containing the three jets,
$
\cos{\psi_3} \equiv {(\vec{p}_3 \times \vec{p}_{B}) \cdot
                              (\vec{p}_4 \times \vec{p}_5)} /
                              {(|\vec{p}_3 \times \vec{p}_{B}|
                               |\vec{p}_4 \times \vec{p}_5|)},
$
where the jets are numbered, 3, 4 and 5 in order
of decreasing energy.
ZEUS has measured the three-jet cross section in the kinematic regime
defined by
$M_{\mbox{\scriptsize 3J}} > 50$~GeV, $|\cos \vartheta_3| < 0.8$ and
$X_3 < 0.95$~\cite{ZEUS_3J}.
(In fact, the events have been further required to have at least two jets with
$E_T^{\mb{jet}} > 6$~GeV and a third jet with
$E_T^{\mb{jet}} > 5$~GeV, however these cuts are largely irrelevant because
high $E_T^{\mb{jet}}$ values are forced by the energy and angular cuts.)
The measured cross section is well described by \oos\ perturbative QCD
calculations both in normalization and in the shapes of the
$M_{\mbox{\scriptsize 3J}}$, $X_i$, $\cos \vartheta_3$ and 
$\psi_3$ distributions.

\begin{figure}[htb]
\begin{center}
\includegraphics[width=.6\textwidth]{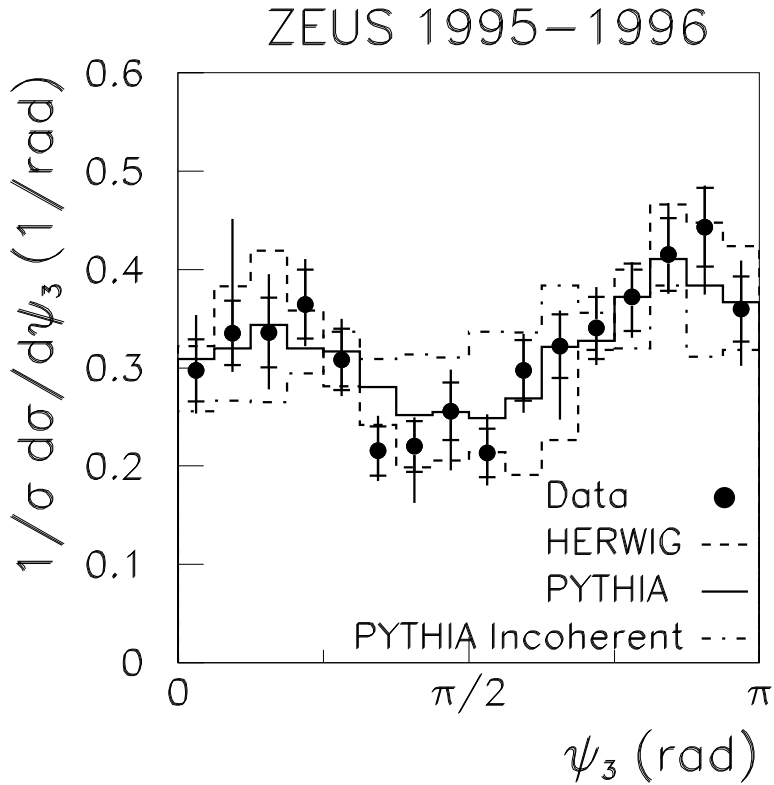}
\end{center}
\caption[]{
The area-normalized distribution of $\psi_3$.
The inner error bar shows the statistical error and the outer error bar
shows the quadratic sum of the statistical and systematic uncertainties.
The solid histogram shows the default PYTHIA prediction.
The dashed and dot-dashed 
histograms show the predictions
of HERWIG and of PYTHIA with colour coherence 
switched off}
\label{fig:ZEUS_psi}
\end{figure}

The measured angular distributions are of particular interest since they
differ markedly from the expectation for three jets distributed evenly
over the available phase space, and are therefore sensitive to the underlying
physical dynamics.  The $\cos \vartheta_3$ distribution, since it is
primarily determined by the distribution of the highest energy jet, is
similar in three jet production to the distribution of $\cos \vartheta^*$
in two jet production.
However, $\psi_3$ is determined by the orientation of the third, or softest,
jet.
For orientations in which the third jet is radiated close to the plane 
defined by the highest energy jet and the beam, $\psi_3 \sim 0$ or $\pi$.
The data are shown in Figure~\ref{fig:ZEUS_psi}.
They indicate that configurations in which the third jet is far from the plane
containing the highest energy jet and the beam ($\psi_3 \sim \pi/2$)
are suppressed.  The dip at $\psi_3 = 0$ and $\pi$ is caused by a loss of
phase space at low angles due to the $E_T^{\mb{jet 3}} > 5$~GeV 
requirement.  Taking this into consideration, the data indicate a strong
tendency for the three-jet plane to lie close to the plane containing the
highest energy jet and the beam.

To aid in the development of a mental picture for three jet production, the
data have been compared to the predictions of the PYTHIA and HERWIG models.
The hard subprocess is included only at leading order in these models
so three-jet events arise from the parton shower phase of the simulation.
Parton showers do a remarkably good job of simulating three-jet production
as is evident from the agreement of the models with the data in the
$\psi_3$ distribution.
Within the PYTHIA model it is possible to switch off the simulation of
QCD colour coherence.  With no simulation of coherence, PYTHIA predicts a
relatively uniform population of the $\psi_3$ distribution, as shown in
Figure~\ref{fig:ZEUS_psi}.
Colour coherence in the parton shower model is required to describe
the observed suppression of large angle radiation.

\section{Jet Substructure}
\label{sec:structure}

The $k_T$ jet-finding algorithm~\cite{kt1,kt2} clusters objects into jets
based on a  distance parameter which is essentially their relative
transverse momentum.  Subjets may be defined within jets by applying the
$k_T$ algorithm to the particles of the jet and counting the subjets as a
function of the resolution parameter, $y_{\mb{cut}}$.  For large values
of  $y_{\mb{cut}}$ there is only one subjet, the jet itself, but as 
$y_{\mb{cut}}$ decreases more and more subjets are resolved until the 
subjet multiplicity equals the multiplicity of particles within the jet.

\begin{figure}[htb]
\begin{center}
\includegraphics[width=.6\textwidth]{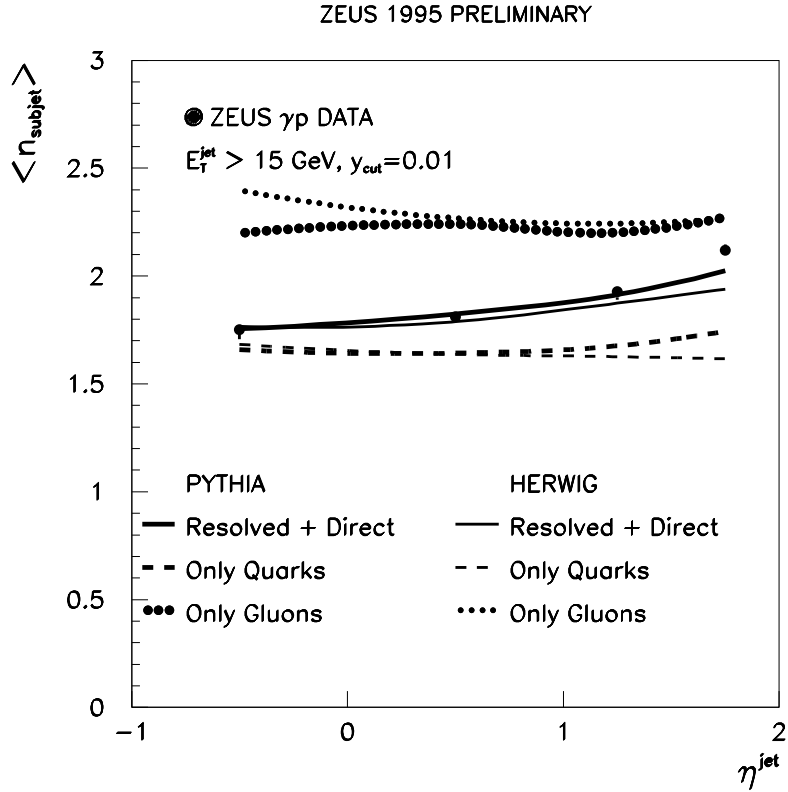}
\end{center}
\caption[]{The mean subjet multiplicity at a fixed value of
$y_{\mb{cut}} = 0.01$ as a function of $\eta^{\tb{jet}}$.  The error
bars show the statistical and systematic uncertainties added in 
quadrature.  For comparison, the predictions of PYTHIA including resolved
plus direct processes for quark jets 
({\it thick dashed line}), gluon jets
({\it thick dotted line}), and all jets ({\it thick solid line}) are 
shown.  The predictions of HERWIG are displayed with thin lines}
\label{fig:ZEUS_subjet}
\end{figure}

The dependence of the average number of subjets, 
$\langle n_{\mb{subjet}} \rangle$,
on  $y_{\mb{cut}}$ has
been measured by the ZEUS collaboration for an inclusive sample of jets
with $E_T^{\mb{jet}} > 15$~GeV~\cite{ZEUS_subjet}.  Both the HERWIG and
PYTHIA models provide a good description of the evolution of
$\langle n_{\mb{subjet}} \rangle$ with $y_{\mb{cut}}$.
Measurements of 
$\langle n_{\mb{subjet}} \rangle$ for $y_{\mb{cut}} = 0.01$ have been
performed in four different jet pseudorapidity regions, as presented in
Figure~\ref{fig:ZEUS_subjet}.
The average number of subjets increases as the jets move toward the
incoming proton direction.  This behaviour is well described by the PYTHIA
and HERWIG models.
In the models the dominant leading order direct process is
$\gamma g \rightarrow q \bar{q}$ while the dominant resolved process is
$q g \rightarrow q g$.  Therefore, relatively more gluon jets are expected
for the more forward boosted resolved photon processes.  Moreover,
the gluon in the $q g \rightarrow q g$ subprocess has a tendency to be the
more forward parton, further increasing the gluonic content of forward jets.
The fundamental expectation of QCD that gluons, which have 
a higher colour charge than quarks, should yield a higher multiplicity of
hadrons, is borne out in the models by a higher average subjet multiplicity
for gluon jets than for quark jets.
Thus the increase of
$\langle n_{\mb{subjet}} \rangle$ with $\eta^{\mb{jet}}$ may be understood to
arise from an increasing admixture of gluon jets in the forward direction.
We look forward to an eventual comparison of this data with perturbative
QCD calculations.

\section{Summary}

Hard photoproduction events have been used in a variety of analyses in order
to further the understanding of the physics of strong interactions.
From studies of the structure of the real photon it has been shown that 
in a leading order interpretation of the data the gluon density rises
as the momentum fraction of the photon accessed in the hard subprocess
decreases.
There is also an indication that the quark densities of the real photon
may be underestimated at high momentum fractions and for high values of the 
probing energy scale.
A global fit of $e \gamma$ and $\gamma p$ measurements should be undertaken to
discover whether a parton density can be found which describes all the data.
Studies of the structure of the virtual photon have illustrated the 
expected suppression in the photon's structure as its lifetime decreases.
Measurements sensitive to the underlying QCD dynamics have shown
that \oos\ perturbative QCD matrix elements,
and models with \os\ matrix elements together with parton showers, are
successful in explaining the mechanisms of high mass dijet and three-jet 
production.
A measurement of jet substructure has shown the sensitivity of jets
produced in hard photoproduction processes to quark and gluon jet
differences and revealed their potential to provide insight into
the physics of hadronization.
Results from photoproduction at HERA have progressed from simple 
manifestations of the photon's hadronic structure, to detailed investigations
of that structure and of QCD in general.
As the yearly luminosity deliverable by HERA continues to increase,
and as the dialogue between theorists and experimentalists is
continuously strengthened,
one may look forward to an even greater variety and quality of physics
result to emerge from the investigation of hard photoproduction at HERA.

\end{document}